\documentclass[sigconf,screen,authorversion,pbalance]{acmart}

\AtBeginDocument{%
  }

\setcopyright{cc}
\copyrightyear{2021}
\acmDOI{10.48550/arXiv.XXXX.XXXXX}

\acmConference{CSE 580: Computing for Social Good}{Spring 2021}{Seattle, WA}
\acmBooktitle{CSE 580: Computing for Social Good}

\begin{document}

\title{Community Cellular Networks Coverage Visualizer}

\author{Chanwut Kittivorawong}
\orcid{0000-0002-2884-2221}
\affiliation{%
  \institution{University of Washington, Seattle}
  \state{Washington}
  \country{USA}}
\email{chanwutk@cs.washington.edu}

\author{Sirapop Theeranantachai}
\orcid{0009-0000-6394-6045}
\affiliation{%
  \institution{University of Washington, Seattle}
  \state{Washington}
  \country{USA}}
\email{stheera@cs.washington.edu}

\author{Nussara Tieanklin}
\orcid{0000-0002-3074-8699}
\affiliation{%
  \institution{University of Washington, Tacoma}
  \state{Washington}
  \country{USA}}
\email{nussara@cs.washington.edu}

\author{Esther Han Beol Jang}
\orcid{0000-0002-5346-2706}
\affiliation{%
  \institution{University of Washington, Seattle}
  \state{Washington}
  \country{USA}}
\email{infrared@cs.washington.edu}

\author{Kurtis Heimerl}
\orcid{0000-0002-0989-5440}
\affiliation{%
  \institution{University of Washington, Seattle}
  \state{Washington}
  \country{USA}}
\email{kheimerl@cs.washington.edu}

\newcommand{\fn}[2]{\footnote{#1: \url{#2}.}}

\newcommand{\figureOverall}{
  \begin{figure}[ht]
    \centering
      \includegraphics[width=\columnwidth]{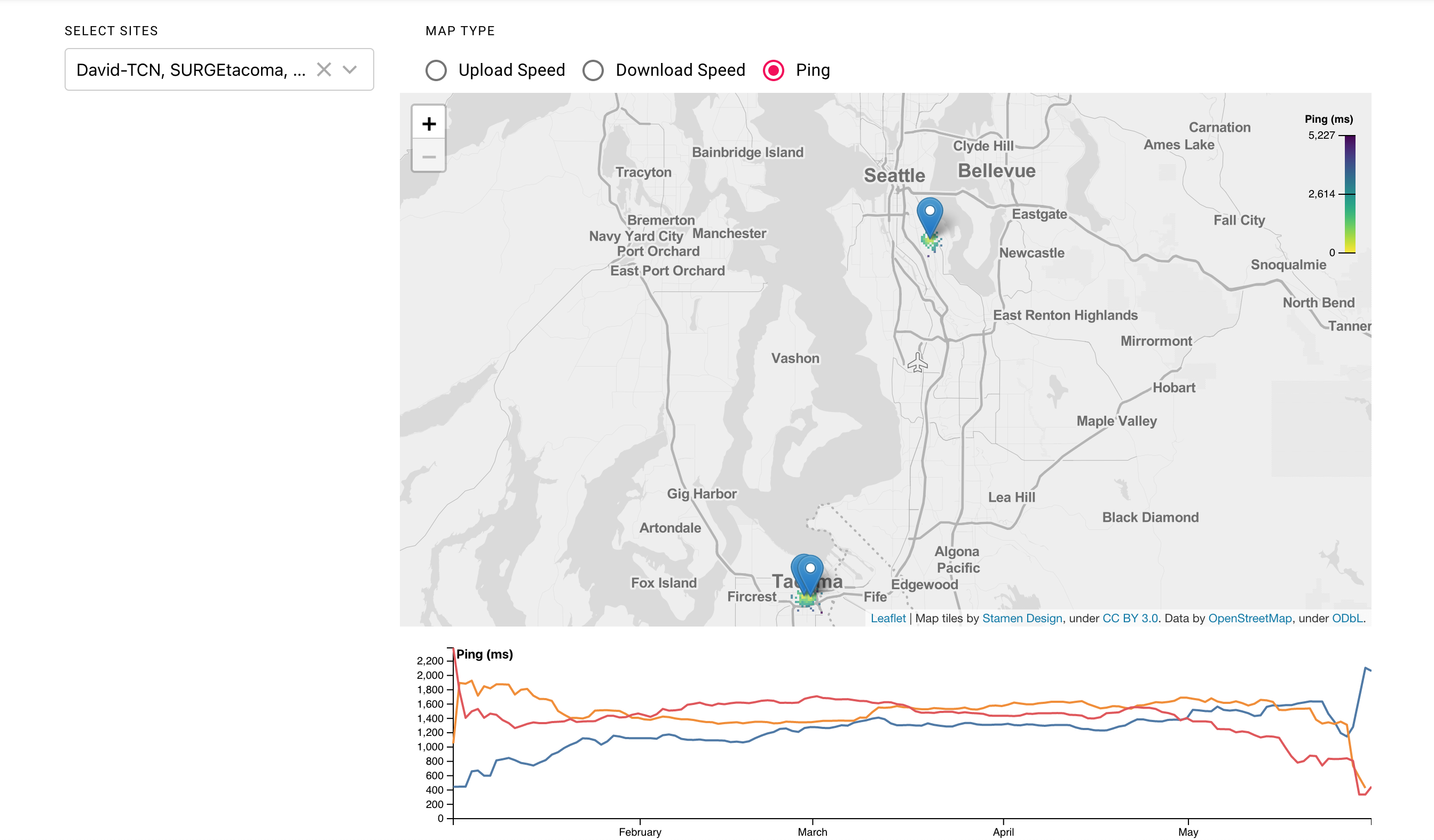}
    \caption{
        The overall interface of the dashboard.
        On the left is the dropdown menu for selecting the base stations.
        On the right, we have
        (1) the radio buttons for selecting measurement type,
        (2) the map for visualizing internet connection quality heatmap, and
        (3) the over-time summary of internet connection quality for each base station.
    }
    \label{fig:overall}
  \end{figure}
}

\newcommand{\figureSite}{
  \begin{figure}[ht]
    \centering
      \includegraphics[width=.47\columnwidth]{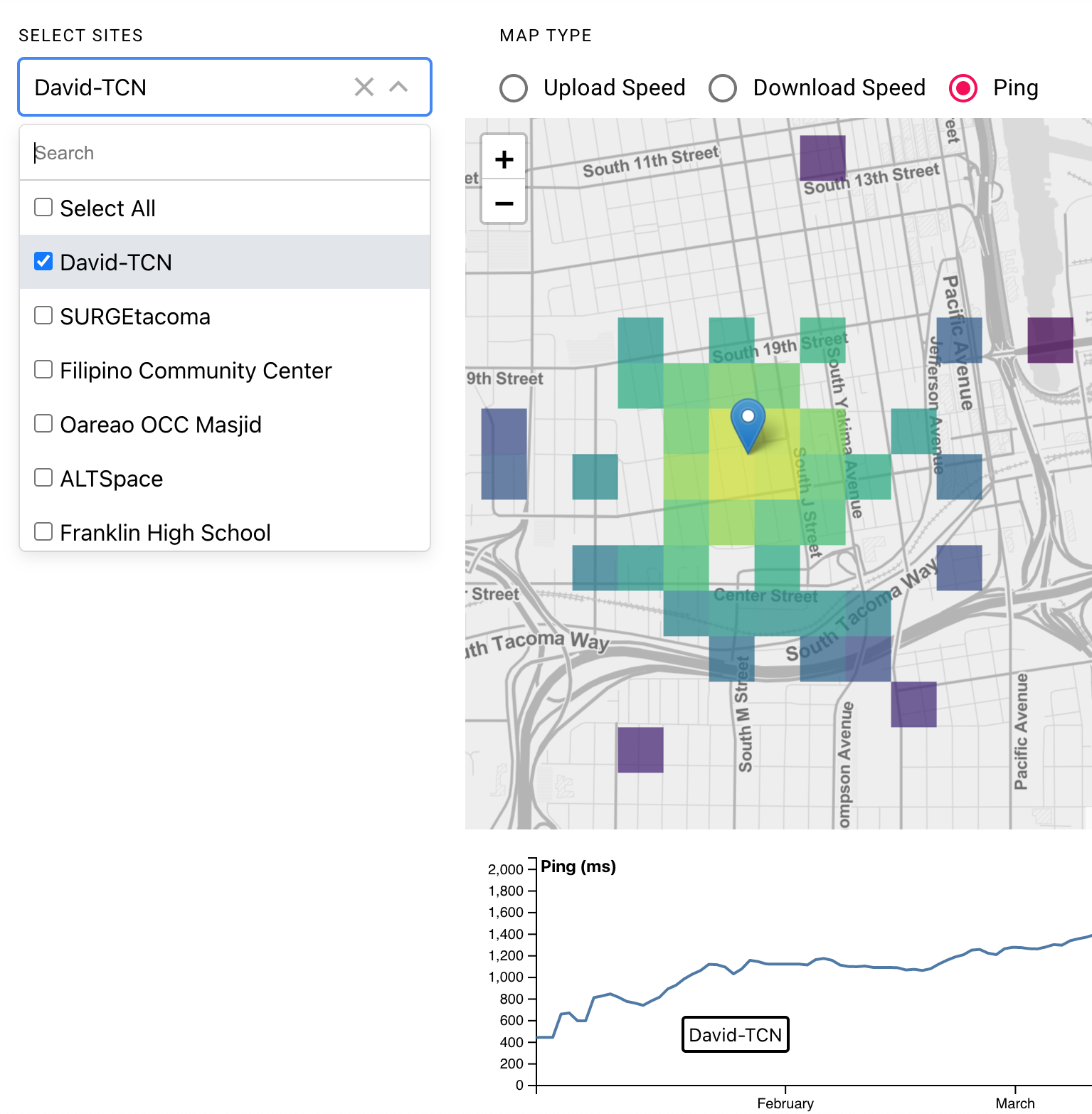}
      \hspace{.04\columnwidth}
      \includegraphics[width=.47\columnwidth]{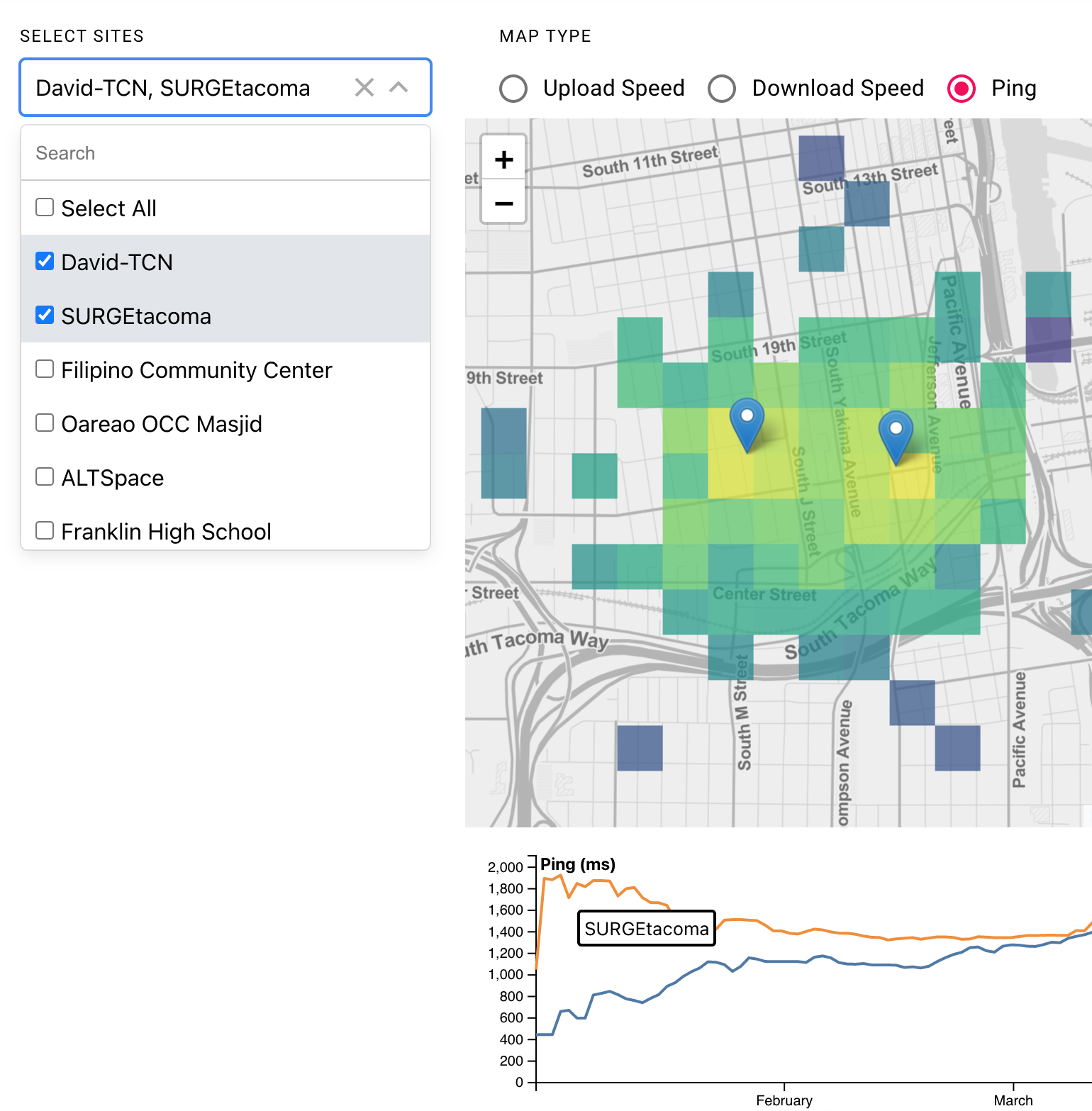}
    \caption{
        Multiple selection of base stations.
        (Left) Viewers only select David-TCN.
        (Right) Viewers additionally select SURGEtacoma.
        The map visualization includes measurements from the devices connected to SURGEtacoma into the heatmap.
        The summary line chart below also adds the line for an overtime average network latency (ping) for SURGEtacoma.
    }
    \label{fig:site}
  \end{figure}
}

\newcommand{\figureTooltip}{
  \begin{figure}[ht]
    \centering
      \includegraphics[width=\columnwidth]{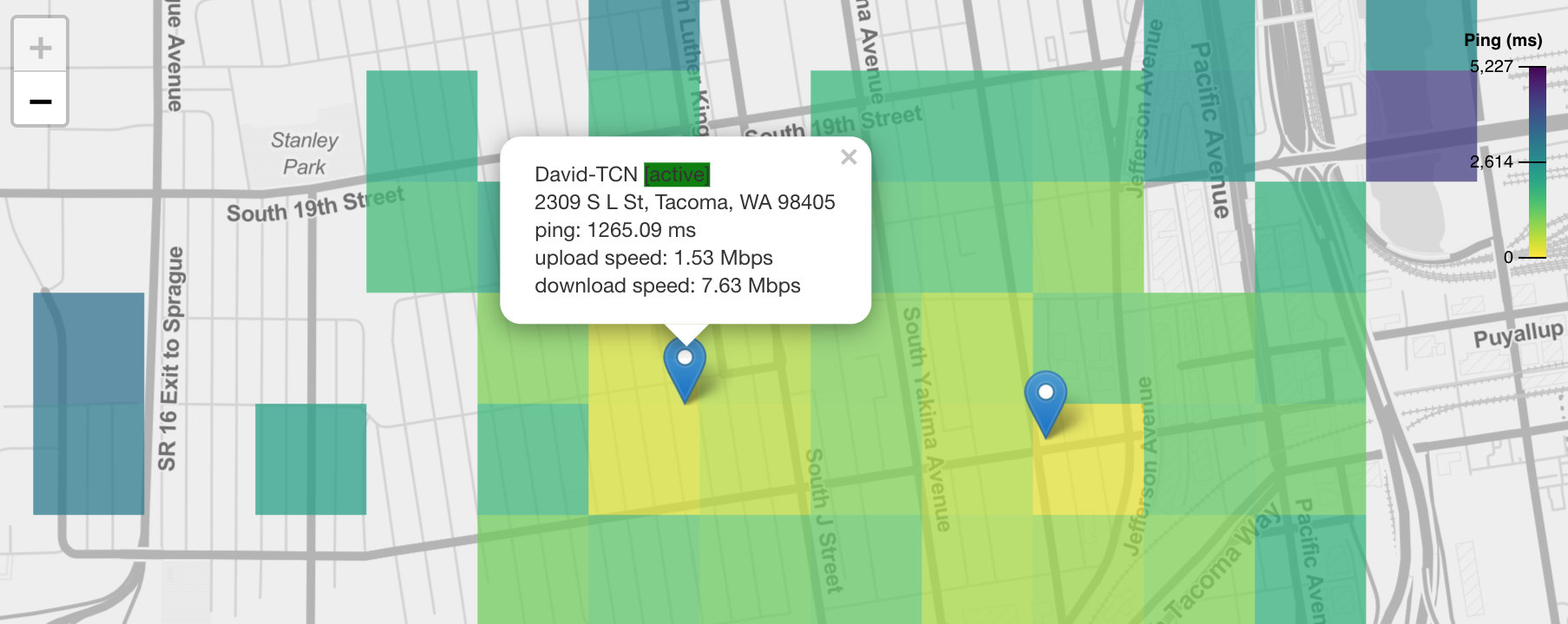}
    \caption{
        We can click a pin to show the tooltip containing
        (1) the name of the base station that the pin represents,
        (2) its status (is the base station currently available),
        (3) its address, and
        (4) the internet connection quality metrics (network latency, upload speed, and download speed).
    }
    \label{fig:tooltip}
  \end{figure}
}

\newcommand{\figureType}{
  \begin{figure}[ht]
    \centering
      \includegraphics[width=.47\columnwidth]{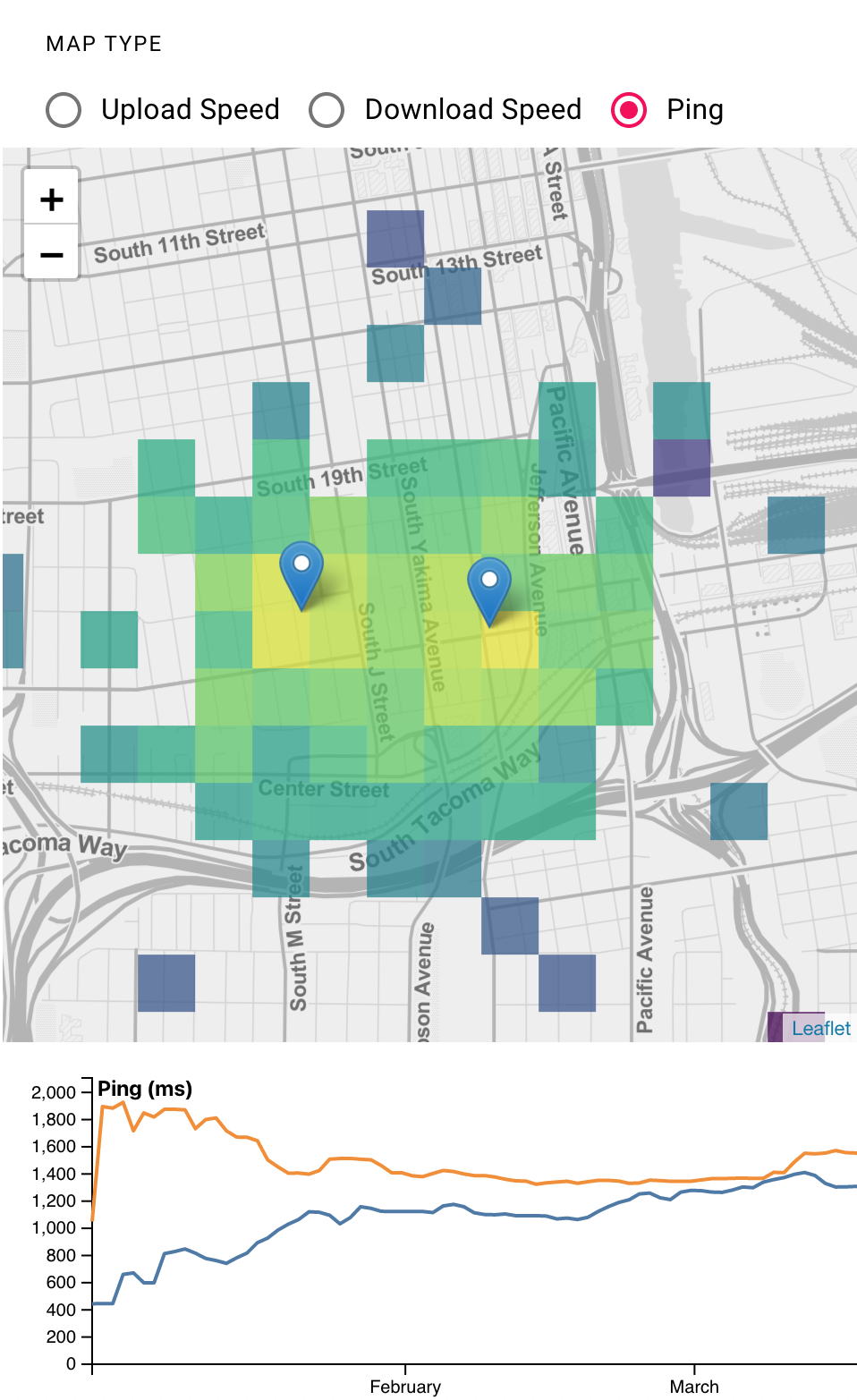}
      \hspace{.04\columnwidth}
      \includegraphics[width=.47\columnwidth]{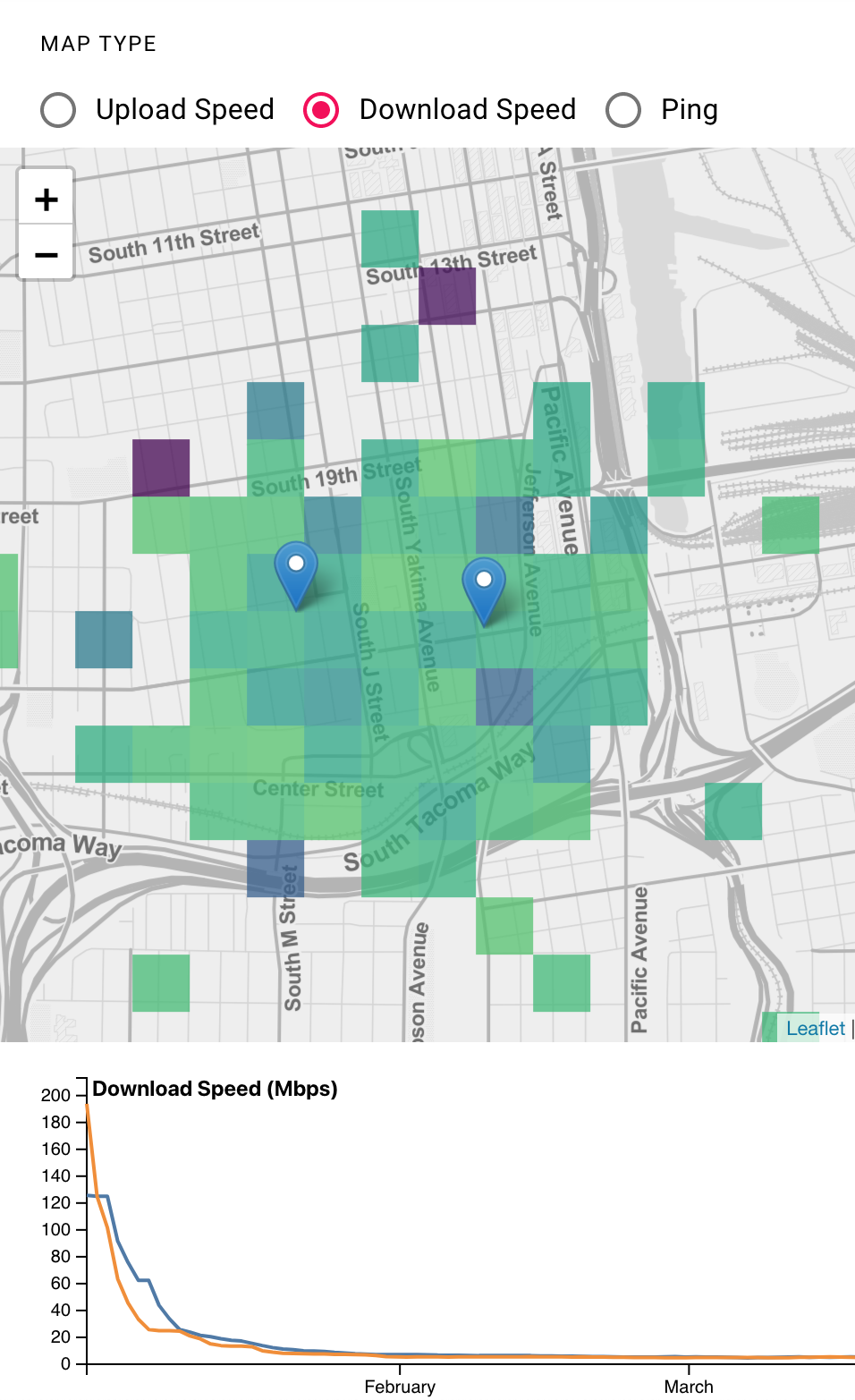}
    \caption{
        (Left) Viewers select network latency (ping).
        The color of each grid box in the heatmap is corresponding to the average network latency in that box.
        The line chart shows the average network latency over time of both base stations.
        (Right) Viewers select Download speed.
        The color of each grid box in the heatmap is corresponding to the average download speed in that box.
        The line chart shows the average download speed over time of both base stations.
    }
    \label{fig:type}
  \end{figure}
}

\newcommand{\figureLoading}{
  \begin{figure}[ht]
    \centering
      \includegraphics[width=\columnwidth]{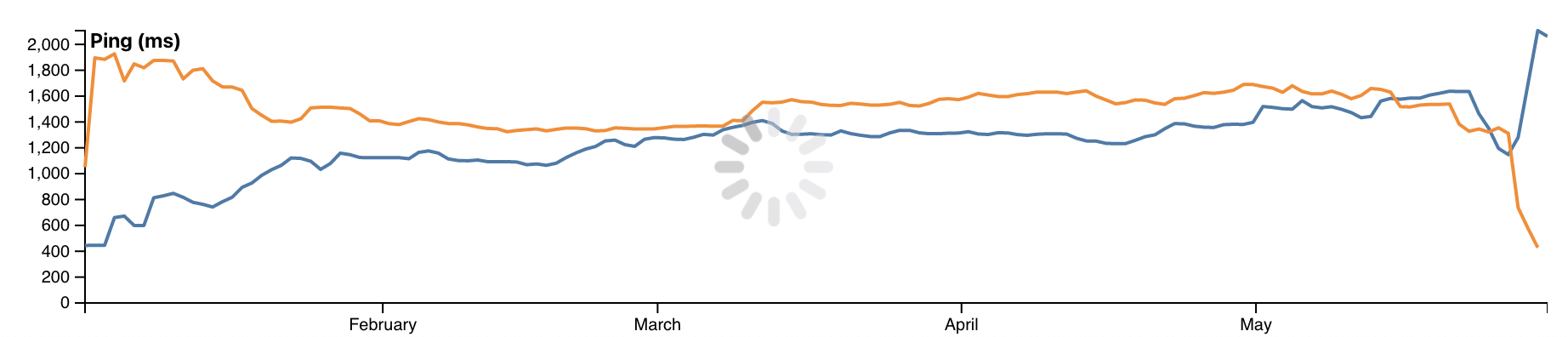}
    \caption{
        A loading sign while the data is being loaded on our summary multi-series line chart.
    }
    \label{fig:loading}
  \end{figure}
}

\newcommand{\figureZoom}{
  \begin{figure}[ht]
    \centering
      \includegraphics[width=.47\columnwidth]{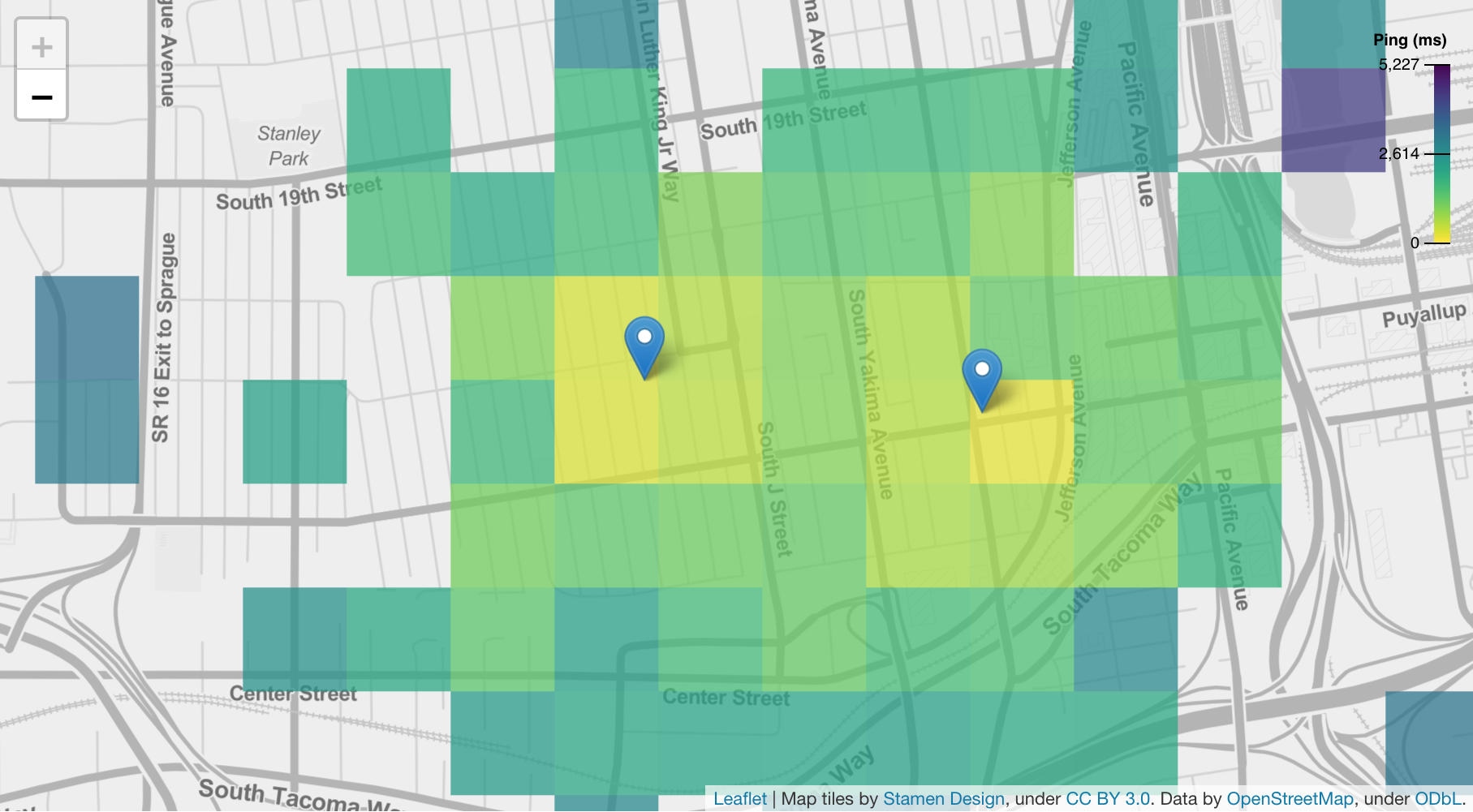}
      \hspace{.04\columnwidth}
      \includegraphics[width=.47\columnwidth]{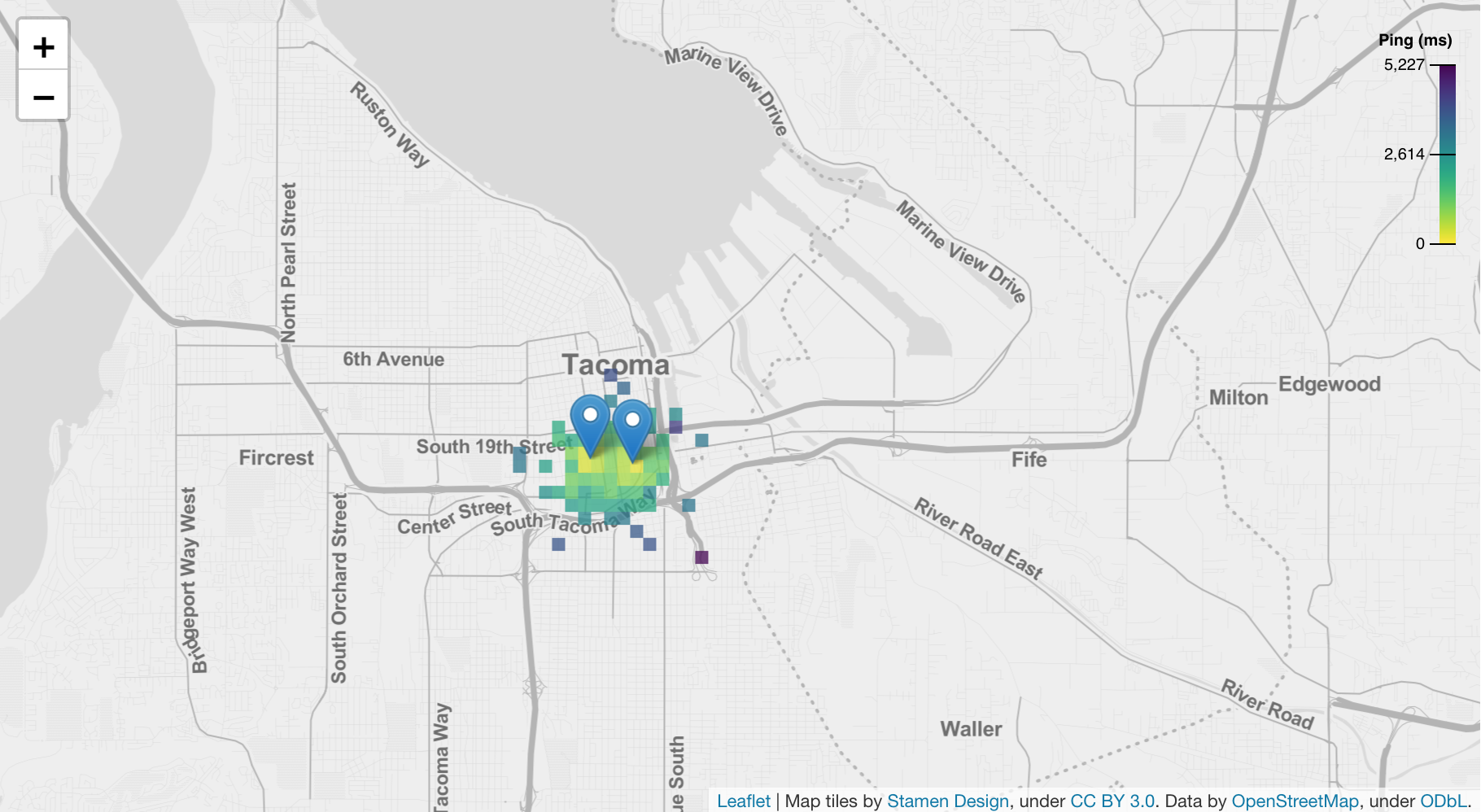}
    \caption{
        Viewers can zoom in (Left) or zoom out (Right) of the map.
    }
    \label{fig:zoom}
  \end{figure}
}

\newcommand{\figureNycMeshI}{
  \begin{figure}[ht]
    \centering
      \includegraphics[width=\linewidth]{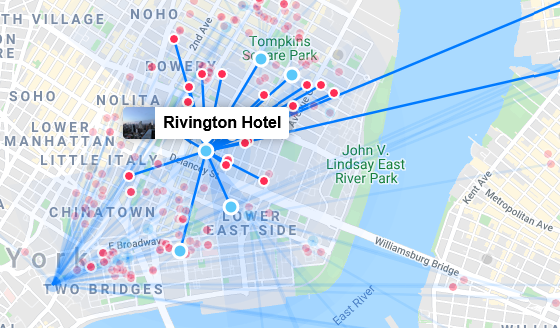}
    \caption{NYC Mesh map overview}
    \label{fig:nycmesh01}
  \end{figure}
}

\newcommand{\figureVizI}{
  \begin{figure}[ht]
    \centering
      \includegraphics[width=\linewidth]{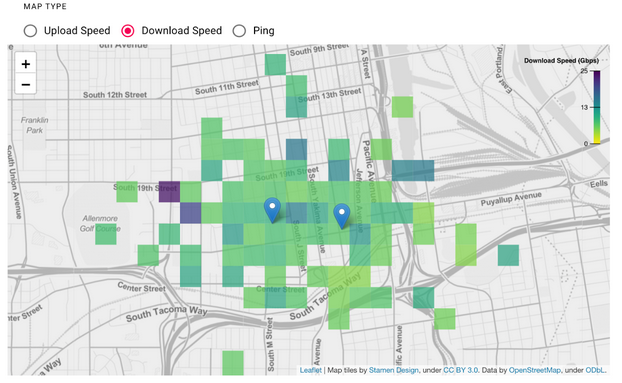}
    \caption{Our map overview}
    \label{fig:viz01}
  \end{figure}
}

\newcommand{\figureNycMeshII}{
  \begin{figure}[ht]
    \centering
      \includegraphics[width=\linewidth]{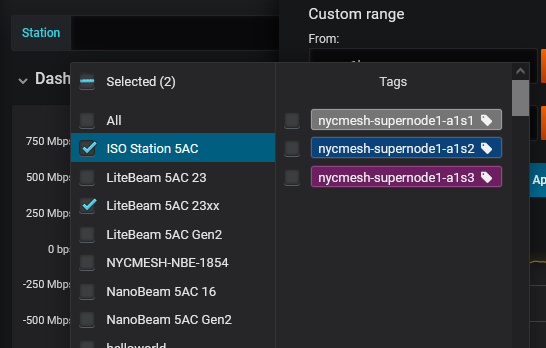}
    \caption{Site selection from NYC Mesh}
    \label{fig:nycmesh02}
  \end{figure}
}

\newcommand{\figureFlentI}{
  \begin{figure}[ht]
    \centering
      \includegraphics[width=\linewidth]{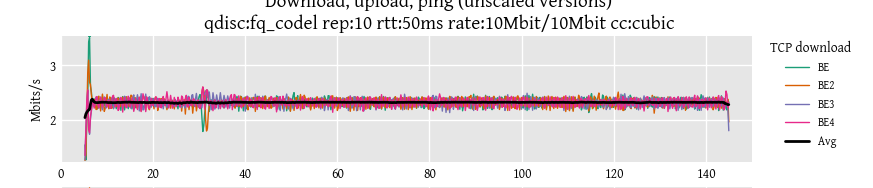}
    \caption{Flent's graphing implementation}
    \label{fig:flent01}
  \end{figure}
}

\begin{abstract}

The community cellular networks volunteers and researchers currently rarely have an access to information about the networks for each site.
This makes it difficult for them to evaluate network performance, identify outrages and downtimes, or even to show the current site locations.
In this paper, we propose the Community Cellular Networks Coverage Visualizer,
a performance dashboard to help reduce the workload of technicians and gain trust from illustrating the reliability of the networks.
The map displays the overall and in-depth performance for each current and future CCNs sites with privacy-focused implementation,
while the multi-series line chart emphasizes on providing the capability of network overtime. 
Not only it will help users identify locations that have stronger and reliable signals nearby,
but our applicaiton will also be an essential tool for volunteers and engineers to determine the optimal locations to install a new site and quickly identify possible network failures.

\end{abstract}




\begin{teaserfigure}
  \includegraphics[width=\textwidth]{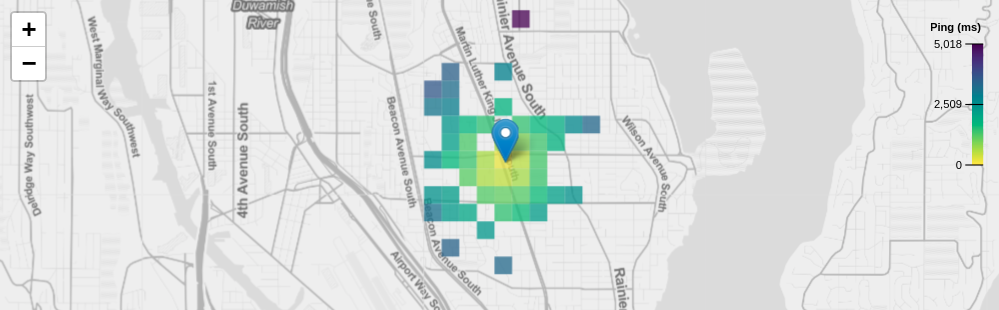}
  \caption{Heatmap of reported network latency from users of Community Cellular Network arround the base station at Filipino Community Center.}
  \label{fig:teaser}
\end{teaserfigure}

\maketitle

\section{Introduction}

According to the UN, as of just last year, 51\% of the global population is still offline, and 7\% live in out-of-coverage areas of any internet provider (ISP), not even mobile phone networks. \cite{itu:net-report} 
People are being disconnected from the rest of the world even in the time where we live, eat, study, or spend the entire day checking the news about the availability of vaccines after the disruption of the Coronavirus (COVID-19) pandemic. 
Over the past year, many research topics have centered around getting underprivileged people access to the necessary resources, including the internet. 
In an attempt to close the digital divide, this project is intended to support the local action research in establishing reliable and robust community cellular networks (CCNs). 

CCNs is an open-source cellular network establishing free or low-cost LTE networks for resource-restricted communities. 
Other than having a lower cost of installation comparing to WiFi network technology,
local grass-root organizations can easily help expand internet access and train people in communities with marketable skills. 
Users will be able to maintain a reliable and fast internet connection without being charged overpriced by ISP.
This will allow the power to build and govern communications networks back into the communities themselves as it is the internet built for the community, by the community.
Yet, network coverage and reliability is the major challenge for every scale of the internet. 
Due to such a wide coverage of at least 2 kilometers radius each site \cite{sevilla:ccn}, it would be difficult for researchers and volunteers to detect the network downtime and locations especially when there will be multiple CCN sites in the near future.

For such a community-operated cellular network, maintaining a stable LTE signal requires performance feedback from the users to determine where the failure occurs. 
To support the future expansion of the network sites,
we propose Community Cellular Network Coverage Visualizer to help maintain reliability and detect any bottlenecks or issues as quickly as possible anywhere you are. 
Our project visualizes the collected data as a map to show the performance within the vicinity of the available sites.
The map locates the areas with the weaker signal transmitted from the individual sites,
allowing us to troubleshoot the network with precision.
Additionally, the dashboard keeps the logs of signals, which can detect the frequency of outages.
These features are intended to mitigate the instability of the CCNs by providing real-time information on the network performance.
In this paper, we will refer to the intended audiences of our application, including CCNs users, volunteers, engineers, researchers and etc. as \emph{viewers}.
In summary, here are three main contributions to the project that we will discuss in this paper:
\begin{enumerate}
    \item
    We designed the visualizations and interactions in the dashboard to helps viewers easily explore any problems with the CCNs.
    
    \item
    We created a robust application with an organized source code for the scalability and extensibility of the application.
    
    \item
    We divide the works of this application into two parts:
    the frontend interface that interacts with the viewers, and the backend server that processes the data.
    This way, we can improve the performance of the frontend interface push all the workload to the backend server.
    We can protect the privacy of CCNs' data by keeping it only in the backend server, preventing any access from the viewers.
\end{enumerate}

\section{Related Work}

\subsection{NYC Mesh}
Before we started our project, we searched for related works that share similar concepts of network visualization to come up with design choices.
The first project is NYC Mesh, made by a dedicated community of volunteers to provide high-speed internet to people in New York City.
This project shares a few traits with the CCNs to adapt features such as an interactive map and comprehensive dashboards to implement our website.

\figureNycMeshI

The map overview of NYC Mesh (see \autoref{fig:nycmesh01}) consists of individual points of connections that form a network.
Since we have a limited amount of data, we need to protect the user's privacy by not directly showing the exact locations of the individuals.
To address the privacy concern, we aggregate points within smaller squares when there are enough points in that area to display the color based on the average ping, upload speed, and download speed.

\figureVizI

Instead of having a line connection between nodes to show what client connects to which site,
we offer the information only for the selected sites to reduce complexity when combined with the color scales.
\autoref{fig:viz01} shows two sites marked by blue pointers.
The colors are filled additively according to the selected sites.

\figureNycMeshII

Another feature we adapted from the NYC Mesh is the site selection dropdown (see \autoref{fig:nycmesh02}).
Notice that the map from NYC Mesh contains complex connections among the nodes and sites;
however, the site selection is only active on the dashboard page.
We improve the design further to get rid of the lines,
which became unnecessary after the addition of site selection and the data aggregation to serve our purpose.
Since the lines are only useful to determine the direct connection between each node and site,
we can safely remove this feature without displaying individual nodes.

In our implementation, we combined both map and dashboard into one page to maintain the consistency of the selected sites.
This dropdown affects the information on the map, graphs, and summary dashboard.
This will allow the combined information to draw a real-time graph for each selected site.
Our performance indicator (see \autoref{fig:loading}) was inspired by Flent (see \autoref{fig:flent01}).

\figureFlentI

We plan to produce three graphs containing the real-time values of network latency, upload speed, and download speed to compare the performances among a set of sites.

\figureLoading

\section{Related Technologies}

Our project is a dashboard visualizing specific information.
We create a single-page web application for the dashboard with the following technologies:

\subsection{React and TypeScript}
We use React~\cite{banks:react} with TypeScript~\cite{bierman:ts} as a framework for building the application.
We choose to use React because it is a high-level language specialized for creating single-page web applications.
With React, we use React component that combines the functionalities of
(1) HTML: defining the contents of the React component,
(2) CSS: defining the styles of the React component, and
(3) TypeScript: defining the behaviors of the React component.
Once created, a React component can be used like an HTML tag in other React components.
This way, we can organize the code that splits into components.
Each component has its well-defined functionality.
Combining them into the web
In the end, all the React Components defined are compiled into HTML, CSS, and JavaScript and can be run normally in any web browser.

We also use TypeScript instead of JavaScript.
JavaScript does not give us any warning or error the program is run.
We have to write code, compile, then run the web application before we see an error if there is any.
In contrast, TypeScript compiler catches warnings and error related to typings by statically type-checking our code as we type.
Our work is visualizing data, and we work with data with multiple shapes and formats.
TypeScript compiler watches and warns us whenever we access fields of data incorrectly or invoke function calls with incorrect parameters or return types.
This work will be deployed for the Local Connectivity Lab\fn{Local Connectivity Lab}{https://seattlecommunitynetwork.org},
and is an opensource project that can be extended in the use of other projects.
It is important to create this project with a tool like TypeScript compiler that can verify the correctness of the program.
This can make the development process slower;
We have to add type annotations or add code to verify typings in addition to the code that drive the core functionallity of the application.
However, it makes the web application much more robust and less error-prone.

\subsection{Leaflet}
Leaflet\fn{Leaflet - a JavaScript library for interactive maps}{https://leafletjs.com}
is a JavaScript library for creating interactive maps.
We are working on geospatial data in this project.
Therefore, we use Leflet for creating an interactive map for visualizing performance measurements of network sites.
In addition, Leaflet integrates well with React.

One challenging aspect of visualizing with geospatial data is the mapping between the latitude and longitude values of a data point to the actual location that the readers are familiar with.
Leaflet provides an easy-to-use tool to project a bird-eye view projection of streets and buildings into a 2 dimensional space.
Then, we can plot our data onto that 2 dimensional space,
by matching the latitude and longitude value of each data point to the latitude and longitude values of the projections.
This way, readers can relate the the location of each data point to the actual location that they are familiar with (street or building names).

\subsection{D3: Data-Driven Documents}
D3~\cite{bostock:d3} is a JavaScript library for creating and manipulating HTML Document Object Model (DOM) based on data.
We use D3 to create the summary chart.
D3 gives a very nice and easy-to-use interface for encoding data into a chart.
We can make a visualization with complex visual elements and interactions,
while keeping the code to be minimal and organized.

\section{Datasets}

Each of the \emph{BaiCells Nova 233 CBRS}\fn{BaiCells Nova 233 CBRS}{https://na.baicells.com/product/Details?id=6b3df7d9-a61d-4286-8771-e5c6c0cfa173} basestation is the access point for home user devices, so these users can access the internet through the base station.
Each device will send back its network measurement data back to us.
Then, we can use these data to analyze the network performance of each community networks base station so that we can make adjustments to improve the network quality if there is any issue.
The device will send us the network measurement data periodically, which includes this information:
\begin{enumerate}
  \item Latitude and longitude of the current location of the device.
  \item Timestamp of the current time that the device sends the data.
  \item Device id, a unique identifier of the device.
  \item The base station  that the device is connecting to.
  \item Network latency of the connection.
  \item Download speed of the connection.
  \item Upload speed of the connection.
\end{enumerate}

However, we do not have the dataset ready for this project, due to the dataset is from another on-going project.
Therefore, we have to generate a mock-up dataset.
This dataset is a mimic of the activities from the three locations that already confirmed to be the base stations for us.
There are 100 devices connected to each base station.
Each of these devices can connect or disconnect its base station any time between February 1, 2021 and July 1, 2021.
While connecting, it generates a network measurement data every 15 minutes.

\section{Methodology}
To show data collected from the devices connected to our base stations,
we create a dashboard with two visualizations (see \autoref{fig:overall}).
\figureOverall

First, we create a cartographic map of Seattle and Tacoma area.
In the map, we overlay a heatmap of a measurement value.
With this map, viewers can see the quality of the internet connection for each area near each base station.
In this project, the quality of the internation connection is measured from three values:
(1) network latency,
(2) upload speed, and
(3) download speed.
In addition, viewers can see the coverage of the internet connection shared from each base station.
Here are the contributions to the dashboard that we want to present in this paper.

Second, we create a line chart for the summary of a measurement of each base station over time.
This is a multi-series line chart.
Each line in the chart represents the mean of measurements from devices connected to a base station.

\subsection{User Interface Design}
For the overall interface, we implemented a dropdown menu on the left of the dashboard.
Viewers can use the dropdown menu to select the base stations that the viewers are interesed in exploring.
Viewers can select multiple base stations in this dropdown menu (see \autoref{fig:site}).
We also implemented radio buttons for selecting the measurement type.
Viewers can select a measurement type that they want to see in the map and in the summary line chart (see \autoref{fig:type}).

\figureSite
\figureType

The main visualization is the cartographic map of Seattle and Tacoma area.
The map has three main components.
\begin{enumerate}
  \item Background projection of streets and buildings and their names.
  \item Pins indicating the locations of community cellular networks base stations.
  \item Heatmap of the quality of the internet connection in the area.
\end{enumerate}

The background projection does not contain any information specific to the data that we work with.
The main purpose of the background projection is for viewers to reference the location of other visual elements in the map with the actual location that they are familiar with.
Therefore, we choose a background projection of streets and buildings with a light grey color scheme.
The grey colors would not be distracting when viewers are trying look at other visual elements.

We add blue pins to indicate the location of the community cellular networks base stations.
The visibility of each pin is controlled by the left dropdown menu.
The ability for toggling the visibility of each pin is important.
It helps the viewers to focus on only the base stations that they are interested in when looking at the heatmap.
In addition, viewers can click each pin (base station) to see a popup annotation (see \autoref{fig:tooltip}).
This annotation contains
(1) the name of the base station,
(2) the location of the base station,
(3) and a list of internet connection qualities of the devices connected to the base station.
With this annotation, the viewers can see the information specific to each pin and a brief summary of internet connection quality of each base station.
\figureTooltip

The main focus of the map is the heatmap that indicates the quality of internet connection in the area.
First, we draw a grid on the map.
Then, we divide the dataset into different groups.
Each group contains all the measurement data points that come from the devices that are in the same grid box.
For each group, we calculate the average of measurement values (either network latency, download speed, or upload speed depending on what viewers choose).
Then, we encode the average value with the grid box's color.
The colorscheme that we use is Viridis\fn{Colormap}{https://bids.github.io/colormap} because it is continuous and readable by visually imparied persons.
Presenting the heatmap in a grid like this gives us and viewers two benefits.
\begin{enumerate}
  \item
  We can pre-compute the value of each grid box.
  So, we do not need to store the whole dataset in the frontend website, making the web application smaller.
  
  \item
  The size of each grid box is about four street blocks.
  Viewers do not have a way to know the actual location of each device that sends the measurement data.
  So, this gives a better privacy to users of the community cellular networks.
\end{enumerate}
Finally, the map can be zoomed-in or zoomed-out to see small or overall details of the heatmap (see \autoref{fig:zoom}).

\figureZoom

Below the map is the summary line chart of each base station.
Each line represents an over-time average of a measurement value from devices connected to the base station.
For example, the right figure on \autoref{fig:site} shows the average network latency (ping) in milliseconds over time.
The blue line shows the average network latency of base station David-TCN.
The orange line shows the average network latency of base station SURGEtacoma.
Viewers can hover over a line to see what base station each line represents (see \autoref{fig:site}).

\subsection{Data Visualization Technique}
As mentioned in the previous sub-section, we can precompute the data for the heatmap to reduce size ofthe frontend website.
We setup our project to have two main servers.
First, we have a frontend server\fn{Github repository for the frontend webpage}{https://github.com/Local-Connectivity-Lab/ccn-coverage-vis} that hosts the website.
This frontend server only contain all functionality of the website written in React and TypeScript without the dataset.
Second, we have a backend server\fn{Github repository for the backend server}{https://github.com/Local-Connectivity-Lab/ccn-coverage-server} that stores the data, preprocesses the data, and sends the data to the frontend website to visualize.

When the frontend website wants to visualize the linechart, it also sends an API request to the backend server.
The request contains
(1) the selected base stations to visualize and
(2) the measurement type to visualize.
For example, the frontend website may send a request to visualize a line chart of download speed for the base stations David-TCN and Filipino Community Center.
Then, the backend server filters in only the measurements from the devices connected to David-TCN or Filipino Community Center.
Then, it finds an average download speed for each hour and for each base station.
Finally, the backend server only sends the aggregated data back to the frontend website to visualize the line chart.

When the frontend website wants to visualize the heatmap, it sends an API request to the backend server.
The request contains
(1) the selected base stations to visualize,
(2) the measurement type to visualize,
(3) the projection parameter from latitude-longitude to x-y in pixels, and
(4) the size of the grid.
For example, the frontend website may send a request to visualize a heatmap of network latency for the base stations David-TCN and SURGEtacoma.
Then, the backend server filters in only the measurements from the devices connected to David-TCN or SURGEtacoma.
Then, it finds an average network latency for each grid box based on the projection parameter and the size of the grid sent from the frontend webpage.
Finally, the backend server only sends the average values of grid boxes back to the frontend website to visualize the heatmap.

The benefits of having a backend server to store and preprocess data are
\begin{enumerate}
  \item
  Viewers have no way to access the dataset, which protects the privacy of the users of the community cellular networks.
  
  \item
  The project can scale larger with more visualizations and larger datasets without significantly increasing the size of the frontend website.
  This make the website accessible to any viewers, even without any powerful computer.
\end{enumerate}

One problem that we find with the frontend-backend model is that the API requests take time,
depending on how powerful is the backend machine.
For this reason, we add a loading sign whenever a viewer interacts with the dashboard, but the backend server has not finished processing the data yet (see \autoref{fig:loading}).
This way, we give a feedback to the viewer that the data is loading, instead of freezing the visualizations.

\section{Conclusion}

In this paper, we present Community Cellular Networks Coverage Visualizer to support the future expansion of the local action research, community cellular networks. 
Our application can help the organizations, volunteers, and researchers to quickly identify and handle any issues that causes unstability of the network.
In addition, the network summary logs with an easy-to-navigate interface allows users to explore the performance overtime gaining trust, reliability, and credibility of the project. 
Our application allows viewers to explore the network coverage and quality with map visualization.
The map has the ability to present overall and in-depth network performance in different areas using a heatmap. 
The multi-series line chart additionally provides network performance logs over time comparing every operating site,
allowing viewers to identify any unusual activities or issues. 
It also supports decision-makings and evaluations, edsorsing future improvements.

More importantly, our application is built specifically to protect user identity as it only presents the aggregated and anynomized reported network data from the server. 
It is impossible to reveal or reverse engineer the location or identity of any CCNs users.
Additionally, preprocessing data collection from a backend server will provide a fast and robust frontend visualization to let viewers explore any issues or questions as soon as they need. 
However, since most of the CCNs sites are in progress of deployment,
we are limited to use only simulated data collection in the current version of the Community Cellular Networks Coverage Visualizer.
Thankfully, as the network is community-operated, we will be able to adapt the crown-sourcing data collecting methodology along with the network expansion in the future work.
The upcoming version of the application will also include the ability of selecting the time range of interest.
With such an insignful and privacy-focused solution,
our application can help reduce the workload of the researchers and volunteers.  
Therefore, Community Cellular Networks Coverage Visualizer is an essential tool in expanding networks more efficiently for the rural and out-of-coverage area,
closing the digital gaps that still continue to grow.

\begin{acks}
We thank Esther Han Beol Jang for the thorough explanations of how community cellular networks work and the supervision through the design process to meet the needs of the viwers.
\end{acks}

\bibliographystyle{ACM-Reference-Format}
\bibliography{main}

\end{document}